\documentclass[prc, preprint, nofootinbib,superscriptaddress,showkeys,showpacs,
   preprintnumbers,amsmath,amssymb,aps,floatfix]{revtex4-1}
\usepackage{graphicx}

\newcommand{\sizefactor}{0.75}

\RequirePackage{lineno}

\usepackage{multirow}
\usepackage{array}
\newcolumntype{L}[1]{>{\raggedright\let\newline\\\arraybackslash\hspace{0pt}}m{#1}}
\newcolumntype{C}[1]{>{\centering\let\newline\\\arraybackslash\hspace{0pt}}m{#1}}
\newcolumntype{R}[1]{>{\raggedleft\let\newline\\\arraybackslash\hspace{0pt}}m{#1}}

\begin{document}

\title{Test of Lorentz invariance in $\beta$ decay of polarized $^{20}\text{Na}$}

\author{A. Sytema}\email[Electronic address: ] {a.sijtema@rug.nl}
\author{J.E.~van~den~Berg}
\author{O.~B\"oll}
\author{D.~Chernowitz}
\author{E.A.~Dijck}
\author{J.O.~Grasdijk}
\author{S.~Hoekstra}
\author{K.~Jungmann}
\author{S.C.~Mathavan}
\author{C. Meinema}
\author{A.~Mohanty}
\author{S.E.~M\"uller}
\altaffiliation{Present address: Helmholtz-Zentrum Dresden-Rossendorf, Germany}
\author{J.P.~Noordmans}
\altaffiliation{Present address: CENTRA, Departamento de F\'isica, Universidade do Algarve, Faro, Portugal}
\author{M.~Nu\~nez Portela}
\author{C.J.G.~Onderwater}
\author{C.~Pijpker}
\author{R.G.E.~Timmermans}
\author{K.K.~Vos}
\author{L.~Willmann}
\author{H.W.~Wilschut}
\affiliation{Van Swinderen Institute for Particle Physics and Gravity, University of Groningen,
 Nijenborgh 4, 9747 AG Groningen, The Netherlands}
\date{\today}

\begin{abstract}
\begin{description}
\item[Background]
Lorentz invariance is key in our understanding of nature, yet relatively few experiments have tested Lorentz invariance in weak interactions.
\item[Purpose]
Obtaining limits on Lorentz-invariance violation in weak interactions, in particular rotational invariance in $\beta$-decay.
\item[Method]
We search for a dependence of the lifetime of $^{20}\text{Na}$ nuclei on the nuclear spin direction.
Such directional dependence would be evidence for Lorentz-invariance violation in weak interactions.
A difference in lifetime between nuclei that are polarized in the east and west direction is searched for.
This difference is maximally sensitive to the rotation of the Earth, while the sidereal dependence is free from most systematic errors.
\item[Results]
The experiment sets a limit of $2\times 10^{-4}$ at 90\,\% C.L. on the amplitude of the sidereal variation of the relative lifetime differences, an improvement by a factor 15 compared to an earlier result.
\item[Conclusions]
No significant violation of Lorentz invariance is found. The result sets limits on parameters of theories describing Lorentz-invariance violation.
\end{description}
\end{abstract}

\maketitle

\section{Introduction}
Lorentz symmetry implies that physical laws do not change under boosts and rotations. 
The theory of General Relativity and the Standard Model of particle physics are both invariant under Lorentz transformations.
One of the frontiers of present-day physics is to unify these theories.
Some of the proposed models allow for Lorentz-Invariance Violation (LIV)~\cite{QG1, QG2, QG3}.
LIV is a manifestation of $CPT$ violation~\cite{Greenberg2002}.
Weak interactions violate the discrete symmetries $C$, $P$, $CP$ and $T$, suggesting the relevance of searches for $CPT$ violation and LIV in weak interactions.
Relatively few searches have been conducted~\cite{tests}. The study of $\beta$-decay can give a unique contribution~\cite{KKVos2015, KV2015a, KV2015}.

We have performed a $\beta$-decay experiment that tests the dependence of the lifetime of nuclei on their absolute orientation.
Such dependence would indicate a violation of rotational invariance, and therefore imply LIV.
The present experiment improves our earlier experiment~\cite{SM2013} in terms of statistical precision and systematic accuracy.
The limit on a sidereal variation of the lifetime has been decreased with one order of magnitude.
This limit can be expressed as limits on the tensor that parametrizes LIV in weak decays~\cite{JN2013}. 
The latter also translates to limits on parameters of the Standard Model Extension (SME)~\cite{SME}.
We will use the theoretical framework of Ref.~\cite{JN2013} to relate our result to those obtained in other experiments.
\section{Principle of the measurement}\label{sec:principle}
Consider a correlation between a preferred direction in absolute space $\hat{N}$ and nuclear spin $\vec{J}$.
For a sample of atoms this correlation can be expressed as
\begin{align}
 \frac{\Gamma}{\Gamma_0} &= 1 + \xi\hat{n} \cdot P\hat{J}\;.\label{termLIV}
\end{align}
Here $\hat{n}$ is the direction $\hat{N}$ transformed to the lab frame.
 $\Gamma$ is the LIV decay rate of polarized atoms and $\Gamma_0$ the Standard Model decay rate.
The average nuclear polarization is $P\hat{J}$.
The magnitude of LIV is $\xi$.

The experiment aims to measure precisely the difference between the lifetimes for opposite polarization directions ($\hat{J}_+ = -\hat{J}_-$), rather than the lifetimes themselves.
This reduces the sensitivity to systematic errors that are common to the two lifetime measurements.
The LIV observable we measure is defined as
\begin{align}\label{eq:LIVsignal}
\Delta_{\text{LIV}} = \frac{\tau^- - \tau^+}{2 \tau}\frac{1}{P_{\text{eff}}} = \xi \hat{n} \cdot \hat{J}_+\;,
\end{align}
with $\tau$ the lifetime taken from literature~\cite{Tilley}. The normalization relative to 2$\tau$ instead off $\tau^- +\tau^+$ is done to avoid dependence on common systematic errors.
$P_{\text{eff}}$ is the effective nuclear polarization.
 It gives the overall sensitivity of the experiment, as discussed in Section~\ref{depolarization}.

For Earth-based experiments
\begin{align}\label{LIVsid}
  \Delta_{\text{LIV}}(t) =&\; \xi N^1 \cos \theta \cos(\Omega t + \phi) + \xi N^2 \cos \theta \sin(\Omega t + \phi) \notag \\
& + \xi N^3 \sin \theta\;,
\end{align}
with $N^{1,2,3}$ orthogonal projections of $\hat{N}$ such that $N^{1,2}$ lie in the equatorial plane. $\theta$ is the angle between the polarization axis and the equatorial plane, $\Omega$ is the Earth's sidereal rotation frequency and $\phi$ is a phase defining $t=0$.
When the polarization direction is in the equatorial plane the sensitivity to LIV amplitudes $\xi N^{1,2}$ will be maximal and the third term in Eq.~(\ref{LIVsid}) is zero.
While Eq.~(\ref{eq:LIVsignal}) has reduced sensitivity to experimental effects that are even in $\hat{J}$ it is sensitive to experimental imperfections which are odd in $\hat{J}$, in particular due to the parity-violating $\beta$-decay.
The latter will cause a systematic offset in Eq.~(\ref{LIVsid}).
The ability to exploit the sidereal dependence to eliminate systematic errors was, indeed, essential to the experiment, and will be discussed in detail in Section~\ref{sec:analysis}. Because of this advantage, our experiment limits LIV at a level close to the statistical limit with a final result of $|A_\text{LIV}|=\xi\sqrt{(N^1)^2 + (N^2)^2}< 2 \times 10^{-4}$.
In our previous experiment the polarization was in the up/down direction. Therefore, the sensitivity to $\xi N^{1,2}$ was reduced with $\cos \theta$ (see Eq.~(\ref{LIVsid})), while the constant term $\xi N^3$ could not be measured because of the aforementioned systematic offset.
\section{Experimental setup and procedures}\label{sec:setup}
\begin{figure*}
	\centering
	\includegraphics[trim = 0 0 0 0, width=\textwidth, clip]{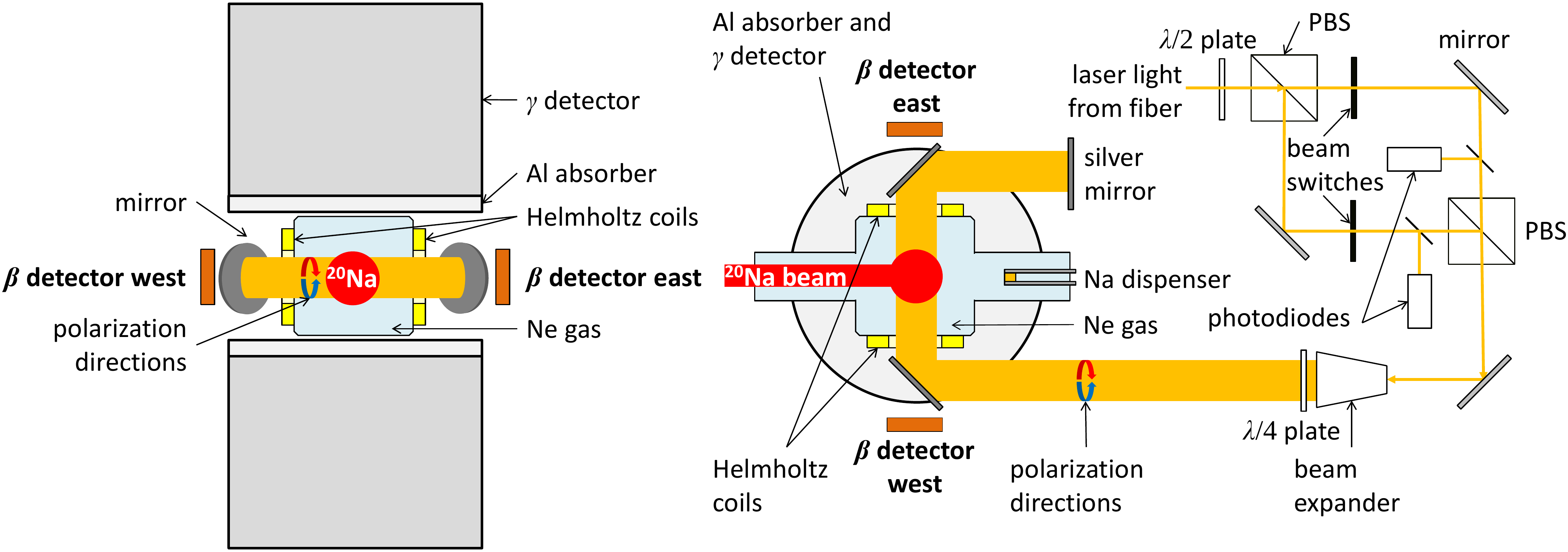}
	\caption{Schematic view of the gas cell, detectors and laser setup. Left: front view; right: top view.}
	\label{setupfig}
\end{figure*}

A $^{20}$Ne beam of 20 MeV/nucleon from the AGOR cyclotron was used to bombard a hydrogen-gas target producing $^{20}$Na at forward angles with similar energy.
The TRI$\mu$P dual separator removes the primary beam giving a beam of radioactive $^{20}\text{Na}$~\cite{Traykov} with $^{19}$Ne as main contaminant.
The $^{20}\text{Na}$ nuclei are stopped in a cell filled with Ne buffer gas, where they can neutralize to atoms~\cite{NIMPolarization}. The $^{19}$Ne stops in the gas-cell window.
Decay rates are measured by $\beta$- and $\gamma$-detectors (see Fig.~\ref{setupfig}).
The $^{20}\text{Na}$ nuclei are polarized in either east ($+$) or west ($-$) direction,
using optical pumping.
The lifetime is extracted from the $\gamma$-decay rates for the two polarization directions.
The experimental method thus assumes that the electromagnetic interaction is Lorentz invariant.
\subsection{Nuclear detection}\label{sec:nucl}
The ground state of $^{20}\text{Na}$ decays with a halflife of 0.45\,s by positron emission.
79\,\% of the decays
are Gamow-Teller transitions to the first excited state of $^{20}$Ne ($E=1.63$\,MeV, $J^\pi = 2^+$).
This state promptly decays with a quadrupole $\gamma$ transition to the $^{20}$Ne ground state.
To be independent of the intrinsic parity-odd emission of the positrons, we use the $\gamma$ ray of 1.63 MeV to signal a decay.
This $\gamma$ ray contributes for more than 99\,\% to the photon spectrum above the annihilation radiation of 0.511 MeV and is therefore the right probe for the selected Gamow-Teller transition.

For $\gamma$-ray detection we placed two large NaI detectors in
the vertical plane. These detectors are placed approximately 75\,mm
away from the center of the buffer-gas cell and have a diameter of 15\,cm. 
The $\gamma$-detection threshold was set at about $1$\,MeV, where the measured spectrum is relatively flat, minimizing the dependence of the count rates on experimental parameters such as gain shifts, drifting offsets and threshold fluctuations.
The contribution of the positrons ($E \leq 11.7\,$MeV) to the $\gamma$ signal has been strongly reduced by placing aluminum absorbers of $20$\,mm thickness in front of the $\gamma$ detectors.
The placement of the $\gamma$ detectors perpendicular to the polarization axis reduces further the asymmetry caused by Bremsstrahlung photons.

For the determination of the polarization $\beta$ detectors are mounted on the east and west side of the gas cell.
The $\Delta E$ (NE-104) scintillator material has $5$\,mm thickness and $44$\,mm diameter.
Low-energy positrons ($E_\beta \lesssim 2\,$MeV) are stopped in the material between the gas cell and $\beta$ detectors.

\begin{figure*}
\centering
\includegraphics{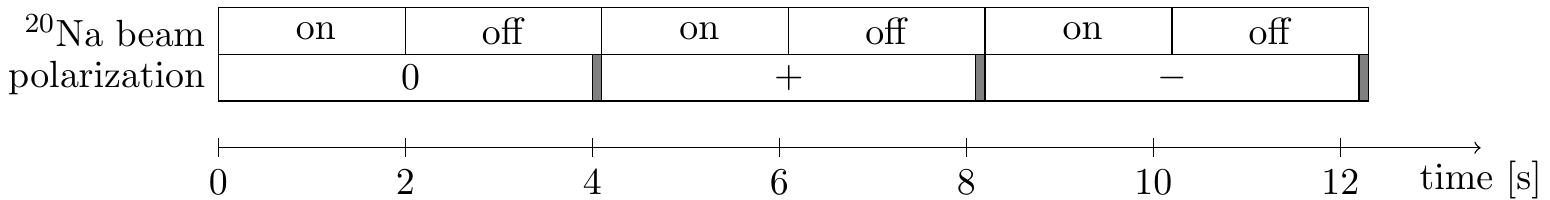}
\caption{\label{fig:switching}The switching scheme of the $^{20}\text{Na}$ beam and the polarization sequence. Gray areas denote the time window for changing the polarization state.}
\end{figure*}

The $^{20}$Na beam was centered in between the main detectors by adjusting the angle of a transmission foil in the incoming $^{20}\text{Na}$ beam, maximizing the count rate in the $\gamma$ detectors.
The $^{20}\text{Na}$ beam was pulsed with beam ``on'' for $2\,$s and ``off'' for $2.1\,$s, respectively, of which the last $0.1\,$s was used for switching the polarization. 
The polarization sequence consisted of three such periods of $4.1\,$s: for unpolarized nuclei and for $\hat{J}_\pm$ polarization (see Fig.~\ref{fig:switching}).
Our data set contains $3\times 10^4$ of such sequences,
with the $\gamma$ detectors having $3.4\times 10^4$ counts on average in a $4.1$\,s period.
\subsection{Polarization}\label{sec:pol}
The polarization of $^{20}$Na is achieved by optical pumping~\cite{Happer}. Details specific to the present experiment are also given in~\cite{NIMPolarization}. 
A solid-state laser system (Toptica TA-SHG pro) provides laser light tuned to the $^2S_{1/2}$ $-$ $^2P_{1/2}$ ($D_1$) transition in $^{20}\text{Na}$ adjusted for the absolute buffer gas pressure of 6.5 bar ($\lambda = 589.782\,\rm{nm}$).
Pressure broadening of about $50\,\rm{GHz}$ mixes the hyperfine levels.
An optical fiber transfers the laser light to the optical table near the gas cell (see Fig.~\ref{setupfig}).
With a polarizing beam splitter (PBS) the laser light is split into two beams with horizontal and vertical polarization. A $\lambda/2$ plate, in front of the PBS, was adjusted to equalize the power of the two laser beams to about $150\,$mW each, in principle sufficient to obtain full nuclear polarization.
With remote-controlled beam stops each laser beam can be blocked.
The beam paths are recombined by a second PBS.
After passing a beam expander, the beam has an approximately Gaussian shape with a full-width-half-maximum of $1.2\,\rm{cm}$.
A $\lambda/4$ plate converts the horizontally and vertically polarized light to circularly polarized light of opposite handedness.
Silver mirrors guide the laser light through the gas cell, passing fused silica windows with a view diameter of $29\,\rm{mm}$.
The windows are surrounded with coils in Helmholtz configuration, providing a magnetic field of about $1.5\times10^{-3}\,\rm{T}$ aligned with the laser beam.

The count rate in the $\beta$ detectors is $R_{\text{E/W}}^\pm\propto 1+ A_{\text{Wu}}\, P\hat{J}_\pm \cdot\vec{\beta}_\text{E/W}$,
with $A_\text{Wu}=1/3$ the $\beta$-asymmetry parameter~\cite{Jackson}.
Here $P\hat{J}_\pm$ refers to the opposite directions of the nuclear polarization with magnitude $P$.
$\vec{\beta}_\text{E/W}$ refers to the velocity relative to the light speed of $\beta$ particles measured in the east (E) and west (W) detector, respectively.
The acceptance in this setup results in $|\langle \hat{J_\pm}\cdot\vec{\beta}_\text{E/W} \rangle| =0.99$.
The $\beta$ asymmetry is obtained from the cross-ratio
\begin{align}\label{eq:Abeta}
A_\beta = \frac{\sqrt{R_{\text{E}}^+R_{\text{W}}^-}-\sqrt{R_{\text{E}}^-R_{\text{W}}^+}}{\sqrt{R_{\text{E}}^+R_{\text{W}}^-}+\sqrt{R_{\text{E}}^-R_{\text{W}}^+}} \approx A_{\text{Wu}} P \;. 
\end{align}
\begin{figure*}[t]
	\centering
	\includegraphics[width=\textwidth]{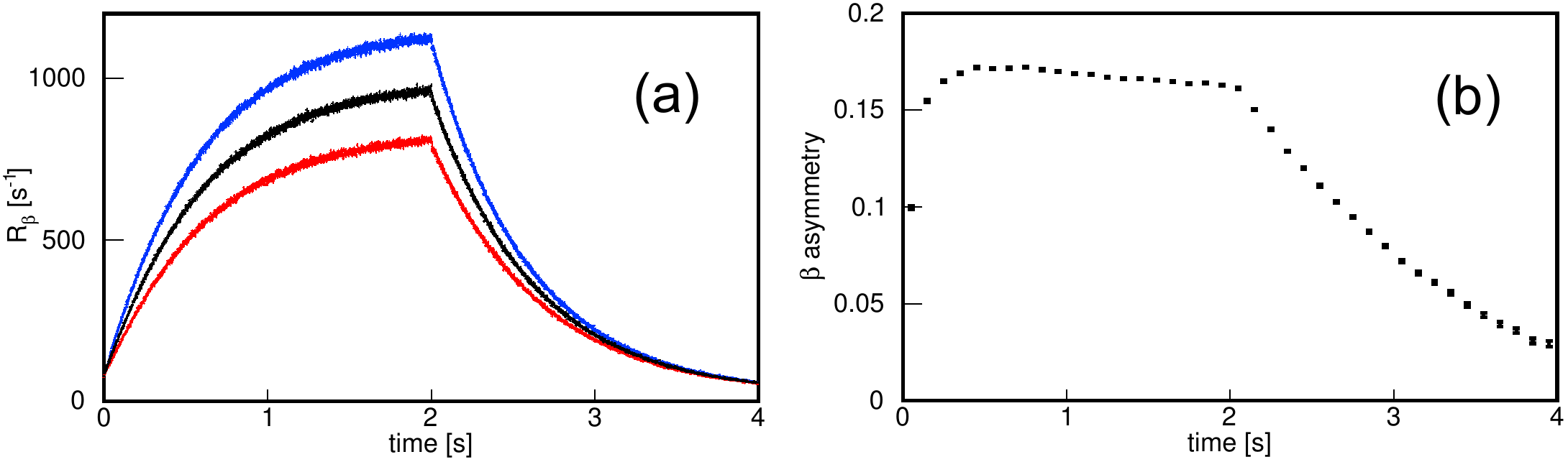}
	\caption{(a) (color online) The instantaneous $\beta$-decay rates averaged over all data of run III for one of the $\beta$-particle detectors. The blue (upper) and red (lower) data points are obtained with opposite polarization.
The black data points (middle) are obtained without polarization. The bin width is 1\,ms.
(b) The experimental $\beta$ asymmetry (Eq.~(\ref{eq:Abeta})). The data points have been binned to 100\,ms reducing statistical scatter.
	}
	\label{fig:betaA}\end{figure*}
	This method for determining $A_\beta$ does not depend on detector acceptance and beam intensity to first order.

The $\beta$ asymmetry as obtained from the weighted average of all data sets is shown in Fig.~\ref{fig:betaA}.
In the first two seconds with beam ``on'' the asymmetry reaches a plateau corresponding to $P= 45\,\%$.
With beam ``off'' the asymmetry appears to decrease exponentially with a lifetime of order one second. 
The loss of polarization can be mainly attributed to molecule formation with residual chemically active reactants.
A detailed account of depolarization mechanisms is given in Ref.~\cite{NIMPolarization}.
This reference also discusses why full polarization is not achieved.
Compared to our previous experiment, the buffer gas pressure has been increased by a factor three to about 6.5 bar.
This reduces both the size of the longitudinal stopping distribution and the diffusion by about a factor three.
At the beginning of the experiment, natural $^{23}\text{Na}$ was evaporated into the buffer gas, which increased the polarization substantially.
The evaporated $^{23}\text{Na}$ is for binding the impurities, that would otherwise bind $^{20}$Na atoms.
Whenever the average polarization dropped during the experiment by about 20\,\%, evaporation of $^{23}\text{Na}$ was repeated.
The polarization improved by a factor two compared to the previous experiment.
\subsection{Additional measurements}\label{additional parameters}
The temperatures of several experimental components were recorded because the expected daily (near-sidereal) variation could introduce a systematic error mimicking a LIV signal. 
The temperature of the gas cell was measured at the position of the $^{23}\text{Na}$ dispenser with a thermocouple.
The other temperatures were measured with platinum resistance thermometers.
The temperature of a metal fence within two meters of the gas cell is indicative of the temperature of the experimental hall.
The temperature of the two large $\gamma$ detectors was measured on the container of the NaI crystal.
The temperatures of the $\beta$ detectors were measured on the metal photomultiplier housing.
Also recorded were the laser-light power for both circular polarization directions using the photodiodes shown in Fig.~\ref{setupfig} and the absolute pressure of the buffer gas.
\section{Analysis}\label{sec:analysis}
Central to the analysis is a multiple-parameter description of the detector count rates to determine $\Delta_\text{LIV}(t)$;
its time dependence should have a period of a sidereal day.
Therefore, variations of experimental conditions on much shorter timescales are reduced by averaging the polarization sequences over a time span of 17 minutes, which we refer to as a \textit{slice}.
Each slice has sufficient counting statistics to perform a multiple-parameter analysis.
Data taking took place during three periods of several days, separated by one month each.
These data sets (labeled I - III) have been analyzed separately using the same procedures.
To perform a blind analysis we randomized the time order of the slices
and determined $\Delta_{\text{LIV}}$ for each slice (Section~\ref{sec:deltaLIV}).
After $\Delta_\text{LIV}$ is determined we apply systematic corrections associated with experimental drifts (Section~\ref{sec:syscor}).
The effective polarization is also accounted for (Section~\ref{depolarization}).
After establishing all analysis procedures, the slices were re-ordered
and analyzed for a possible sidereal variation.
\subsection{Determining $\Delta_\text{LIV}$}\label{sec:deltaLIV}
\begin{figure}
\centering
\includegraphics[width=.45 \textwidth]{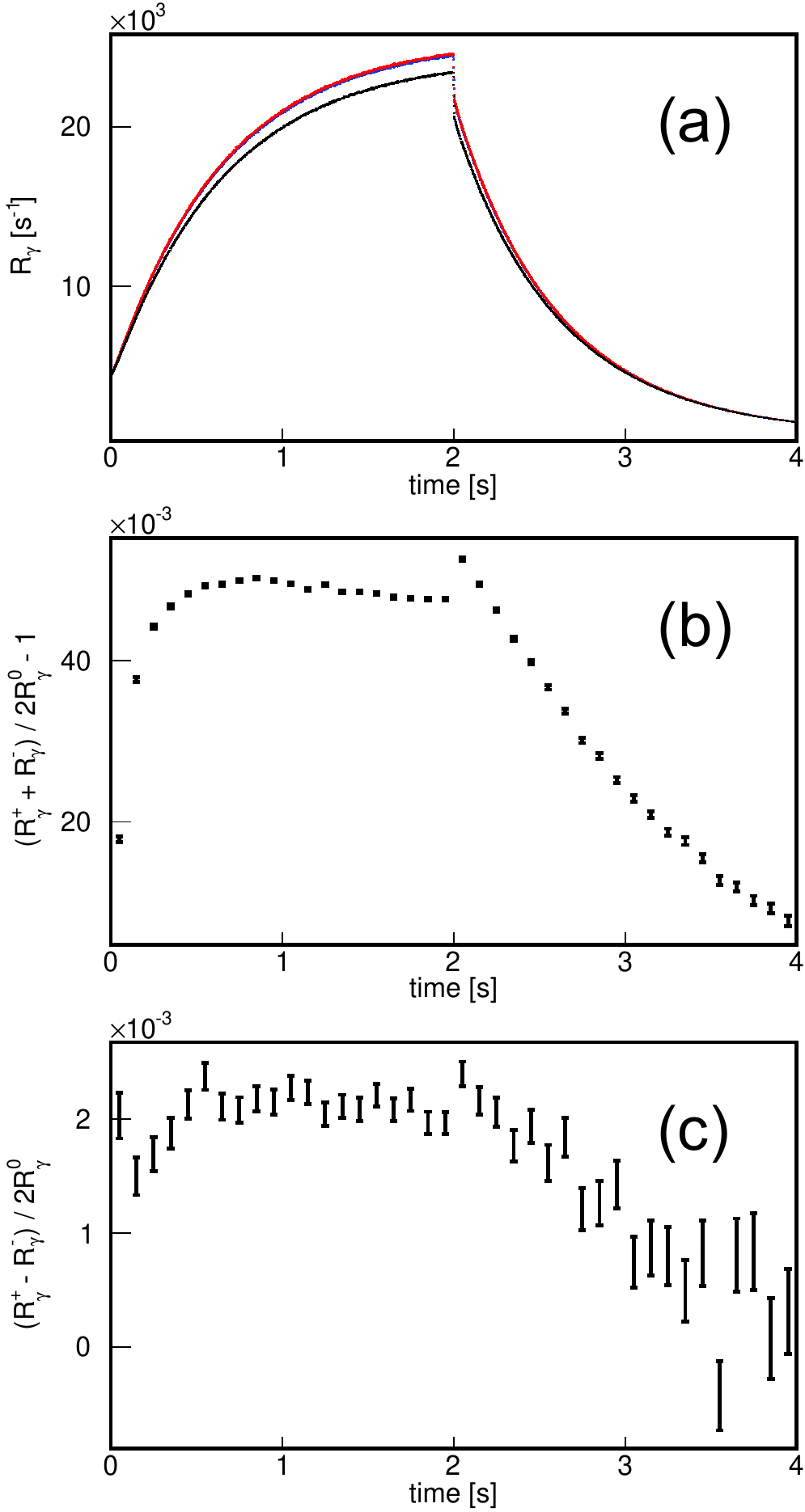}
\caption{Experimental $\gamma$-rate data averaged over run III. \\(a) (color online) The red and blue data points are instantaneous $\gamma$ rates measured in 1\,ms obtained for opposite polarization directions. The blue points are mostly invisible, because they lie under the red points. The sudden drop in count rate at $T=2$\;s is a result of a count rate background which is only present when the production beam is on. \\(b) The presence of the term $P(t)K$ in Eq.~(\ref{eq:countrate}) as seen from the difference between the $\gamma$ rates with and without polarization. The data follow the $\beta$ asymmetry in Fig.~\ref{fig:betaA}.
The jump at $T=2$\;s is again due to the beam related background. The data were binned to 100\,ms.\\(c) A small parity-odd dependence $P(t)L$ in Eq.~(\ref{eq:countrate}) can be seen from the difference between the $\gamma$ rates obtained with opposite polarization. It has an instrumental origin. The data were binned to 100\,ms.}\label{fig:gammarates}
\end{figure}
The $\gamma$-decay rates (see Fig.~\ref{fig:gammarates}) were modeled in detail.
A single $4.1\,$s period adds
\begin{align}\label{eq:basicrate}
R_\gamma(t)=\begin{cases}  A(1-e^{-t/\tau}) & t < T\\
A(e^{T/\tau} - 1)e^{-t/\tau} & t \geq T
 \end{cases}
\end{align}
to the total decay rate, with $A$ the normalization parameter, $\tau$ the lifetime parameter and $T=2\,\rm{s}$ ``on'' time of the beam. To include the
contributions from all previous beam ``on'' periods Eq.~(\ref{eq:basicrate}) is modified. The resulting expression is given in Appendix A.
During the beam ``on'' period, prompt $\gamma$-rays from the production target and primary beam stop added to the detector rate.
This rate (parameter $A_\text{on}$) was modeled with a block function following the time structure of the beam and was typically 15\,\% of the rate maximum. 
Long lifetime components can be modeled as a constant background ($A_\text{bg}$), typically 5-6\,\% of the maximal rate. These two background parameters are independent of polarization.

We include two polarization dependencies in the $\gamma$ count rates as
\begin{align}\label{eq:countrate}
 R_\gamma^\pm(t) = R_\gamma^0(t)\left[ 1 + P(t) (K \pm L )\right]\;.
\end{align}
For each detector $R_\gamma^0(t)$ is the count rate for no polarization and $R_\gamma^{\pm}(t)$ is the count rate for $\hat{J}_\pm$ polarization.
$P(t)$ is parametrized with a polarization rate $\tau_\text{pol}^{-1}$, a polarization-decay rate $\tau_\text{depol}^{-1}$ and a normalization $P_0$. The rate $\tau_\text{pol}^{-1}$ is fixed for each run, the latter two parameters are left free. 
The $K$ and $L$ parameters determine the strength
of parity even and odd decay-rate contributions, respectively.
The rate-enhancement $K$ results from the emission pattern for the quadrupole $\gamma$ transition of the first excited state of the daughter nucleus $^{20}\text{Ne}$ to the ground state.
The quadrupole pattern in the decay of fully polarized nuclei $2^+\stackrel{\text{GT~}\beta}{\longrightarrow} 2^+ \stackrel{\text{E2~}\gamma}{\longrightarrow} 0^+$ has an enhancement perpendicular to the polarization direction of $25\,\%$ compared to isotropic emission~\cite{Tolhoek}. This enhancement is
 $10\,\%$ when integrating over the acceptance of the $\gamma$ detectors.
The enhancement can be seen in Fig.~\ref{fig:gammarates}b where $\gamma$ rates with and without polarization are compared. It follows the polarization
with a plateau value for $K\,P(t)$ of $4.5\,\%$ consistent with the observed maximum polarization of $45\,\%$.
$K$ is left free for both detectors separately ($K_1$, $K_2$) in the fitting procedure.
\begin{table}
 \begin{tabular}{|c|c|c|c|}
\hline
\hline
 \multirow{2}{20mm}{parameter} & \multicolumn{3}{c|}{detector}\\
\cline{2-4}
&$\gamma_1$ & $\gamma_2$ & $\gamma_3$ \\
\hline
\hline
rate normalization & $A_1$& $A_2$& $A_3$\\\hline
overall lifetime & \multicolumn{3}{c|}{$\tau_0$ } \\\hline
lifetime difference& \multicolumn{2}{c|}{$\Delta\tau$ }&- \\\hline
prompt background & $A_\text{on,1}$ & $A_\text{on,2}$ & $A_\text{on,3}$  \\\hline
background & $A_\text{bg,1}$& $A_\text{bg,2}$& -\\\hline
rate normalization $^{19}$Ne &-&-& $A_\text{$^{19}$Ne}$\\\hline
quadrupole parameter & $K_1$ & $K_2$ & -\\\hline
asymmetry parameter  ($\star$)&\multicolumn{2}{c|}{$L$}  & -\\\hline
rate-dependence & $\alpha_1$ & $\alpha_2$&$\alpha_3$ \\\hline
normalization polarization & \multicolumn{2}{c|}{$P_0$} &-\\\hline
polarization rate ($\star$) &\multicolumn{2}{c|}{ $\tau_\text{pol}^{-1}$}&- \\\hline
polarization decay rate &\multicolumn{2}{c|}{ $\tau_\text{depol}^{-1}$}&- \\\hline
\hline
 \end{tabular}\caption{Overview of parameters to describe  the various detector count rates and the polarization. 
 The parameters are determined per 17 minute data slice, except for the parameters $L$ and $\tau_\text{pol}^{-1}$ ($\star$) which are fixed for each of the three runs (see text).}\label{tab:parameters}
\end{table}

The parameter $L$ describes a parity-odd dependence in the $\gamma$-rates and should be absent in an ideal experiment. However, in Fig.~\ref{fig:gammarates}c it can be seen that such dependence exists and reaches a value $L\,P(t)\approx 0.0022$, \textit{i.e.} $L\approx 0.005$.
This can be attributed to the asymmetric distribution of matter around the setup, where positrons annihilate and add to the $\gamma$ signal. Another source of asymmetry could be imperfect balancing of the opposite polarizations.
In practice we use $L$ as the final tuning parameter making the time average $\langle\Delta_\text{LIV}(t)\rangle=0$. The direct connection with the actual parameter is thereby lost and other aspects of the setup asymmetry enter. The largest value of $L=0.015$ is three times larger than expected on basis of the count-rate distributions alone. Systematic dependencies and errors in $\Delta_\text{LIV}(t)$ are discussed in Section~\ref{sec:deltaLIV}.

Most of the $^{19}\text{Ne}$ is stopped in the entrance foil to the gas cell.
No contribution with a $^{19}\text{Ne}$ lifetime was found in the $\beta$ detectors.
To monitor the running conditions an auxiliary detector with threshold below $511\,\rm{keV}$ was placed close to the entrance foil. About 72\,\% of the count rate of the auxiliary $\gamma$-detector consists of $^{19}\text{Ne}$ decays with $T_{1/2}=17$\,s. For this detector, $K=L=0$, also $\Delta \tau=0$.
 Including the auxiliary detector in the fitting routine improved the overall $\chi^2$ for all data slices. 

Finally, taking into account the effects of pile-up, dead time and rate-dependent gain, a term quadratic in count rate was added with a proportionality $\alpha$. The apparent maximal pile-up was typically 5\,\%. An overview of all parameters is given in Table~\ref{tab:parameters}.
We use a $\chi^2$ minimization to fit the set of parameters, except $L$ and $\tau_\text{pol}$, simultaneously to the nine count-rate spectra $R_{\gamma_{1,2,3}}^{0,+,-}(t)$ where $\Delta\tau\equiv (\tau^- - \tau^+)/2$ of the three detectors and three polarization states for each slice of data of 17 minutes.

\begin{figure}
\includegraphics[width=\sizefactor \linewidth]{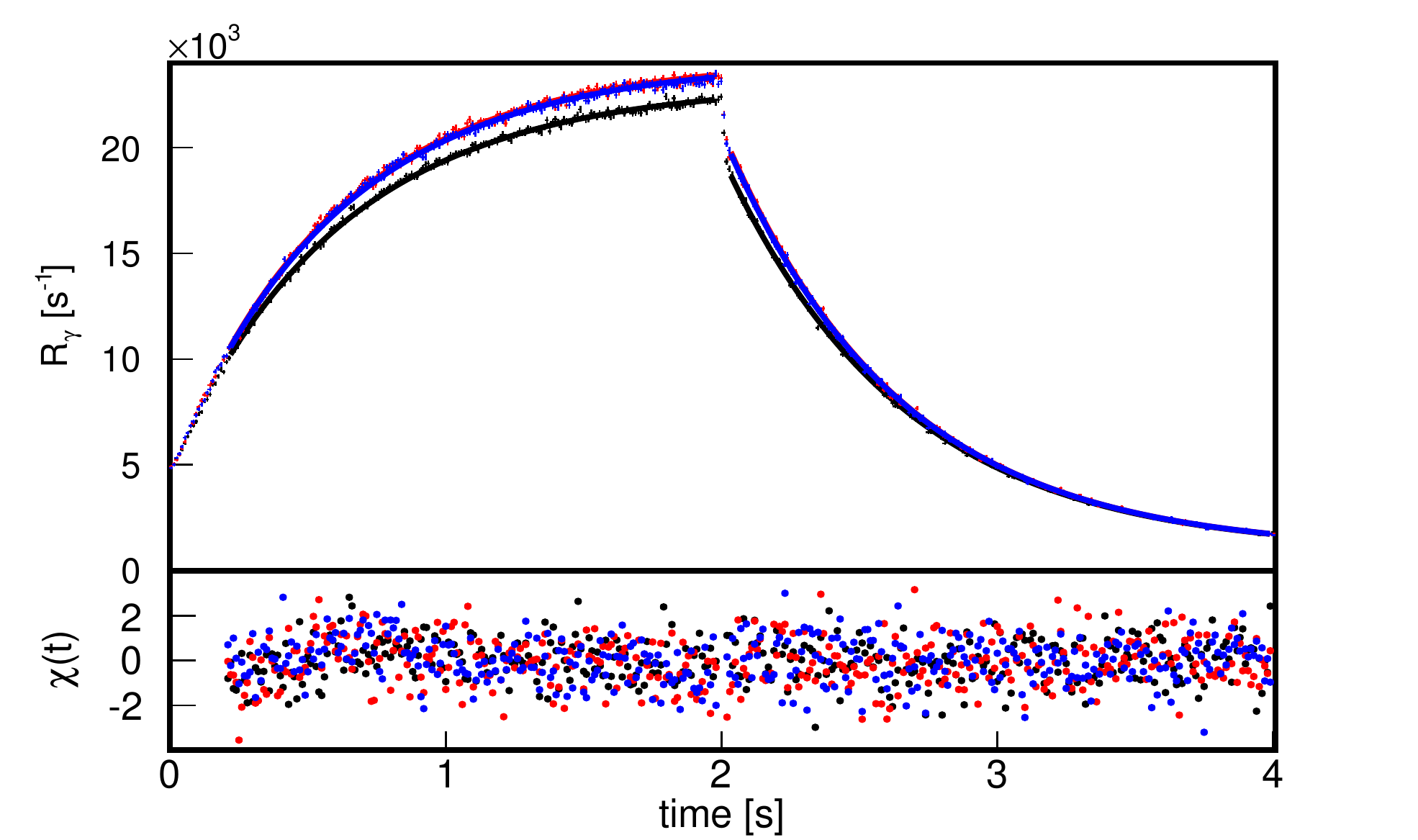}	
	\caption{\label{fitgamma}A typical fit to a $\gamma$-rate spectrum of a single slice of 17 minute data. The fitted lines cover the data, measured as instantaneous rate in 10\,ms.
The overlapping blue and red (upper) data points are obtained with opposite polarization. The black (lower) data points are obtained without polarization.
The time region (0 - 0.21)$\,$s where the polarization may not have lost memory of the previous 4.1\,s polarization period is excluded from the fit. In addition, the region (1.98 - 2.02)$\,$s where the beam is switched from ``on'' to ``off'' is excluded. The bottom graph shows the residuals $\chi=(\text{data}-\text{fit})/\sigma_\text{data}$.} \label{fig:fits} 
\end{figure}
A typical fit of $\gamma$-rate spectra is shown in Fig.~\ref{fitgamma}. The residuals of the fit show the appropriate statistical scatter.
The lifetime $\tau^0$ was found to be about $4\,\%$ smaller than the literature value of 0.45\,s.

To obtain an initial value for $\Delta_\text{LIV}$ we use $\Delta \tau / (\tau P(t=2))$, this value is too large in view of depolarization, as we will argue in Section~\ref{sec:sensitivity}, where we also discuss how to correct for this effect.
\subsection{Systematic corrections to $\Delta_\text{LIV}$}\label{sec:syscor}
The parametrization of the $\gamma$ rate
does not explicitly account for drifts in the experimental equipment.
Therefore, $\Delta_\text{LIV}$ still depends in an intricate way on temperature, cell pressure, {\it etc}.
These can have day-night dependencies that appear as sidereal variation in $\Delta_\text{LIV}$. Their values were recorded in parallel with the data taking. In the following we refer to these as external parameters.
We use average values for each 17 minute slice.
 
We consider the correlations between $\Delta_\text{LIV}^i$ and parameters $p_j^i$, where the index $i$ refers to the individual 17 minute slices. $p_j^i$ is the corresponding average value of an external parameter $j$ such as temperature and pressure. Using vector notation for the data set $\{ i\}$, the correlation is given by $D_j=\langle{\bf\Delta_\text{LIV}\cdot p}_j\rangle$. 
This value can be established without unblinding the data.
\begin{figure}
\includegraphics[width=\sizefactor \linewidth]{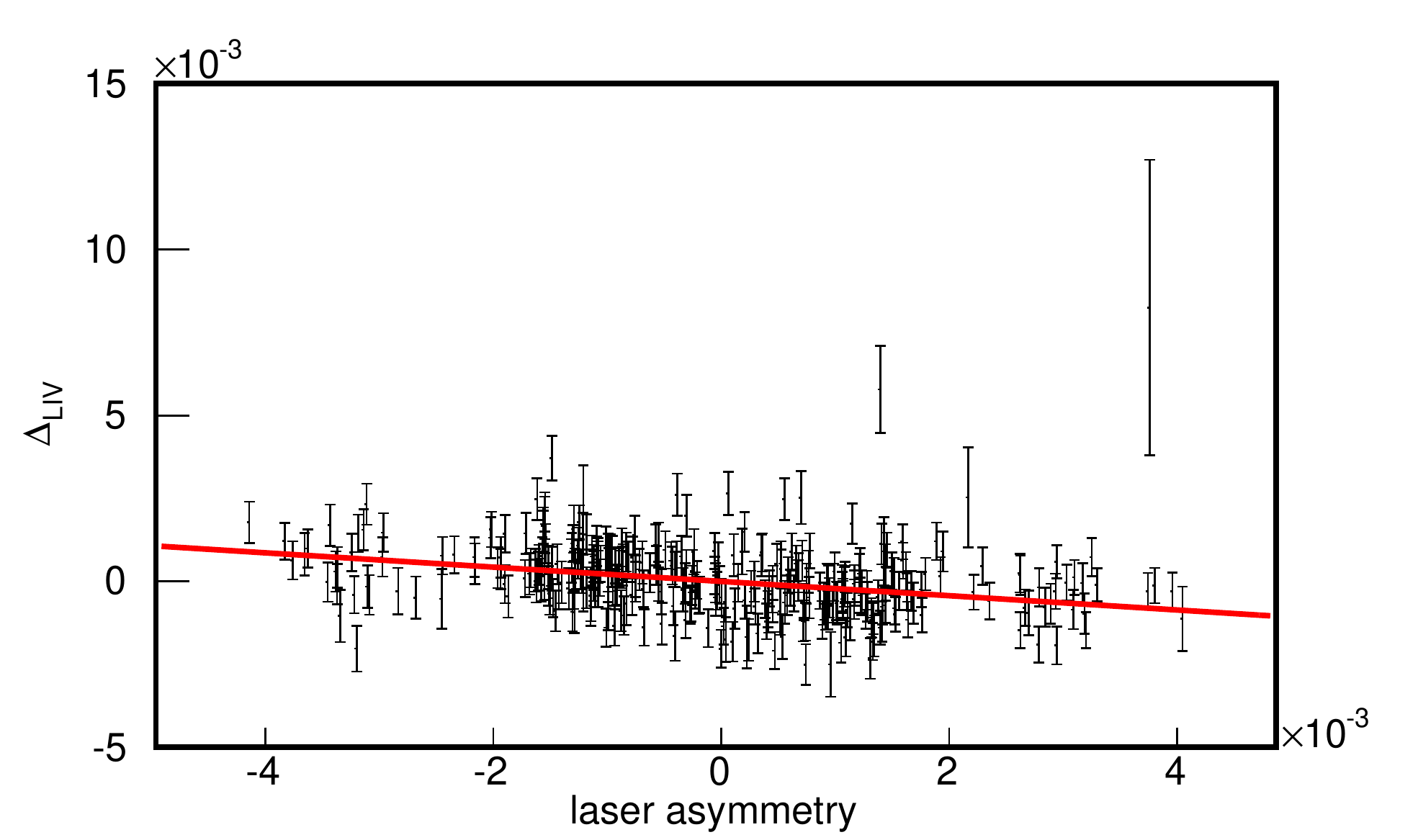}
\caption{The dependence of $\Delta_\text{LIV}$ on the asymmetry of the power of the laser for opposite polarizations (defined as fraction of the difference over the sum). This indicates that the polarizations were not completely balanced throughout the experiment. The laser asymmetry itself was found to be temperature dependent.
Therefore, it is also a measure of the temperature of the experimental hall.}\label{fig:laserasymmetry}
\end{figure} 
The most relevant correlation was found to be the asymmetry in laser power for both polarizations which is shown in Fig.~\ref{fig:laserasymmetry}. The dependence on the parameters $j$ is removed from the data by redefining $\bf\Delta_{\text{LIV}}$ as ${\bf\Delta'_{\text{LIV}}=\Delta_{\text{LIV}}}-D_j\;{\bf p}_j$. We first remove the dependence on the laser asymmetry ($j=las$). Because most parameter drifts are temperature driven we also make sure to remove the correlations $\langle {\bf p}_{las}\cdot{\bf p}_k\rangle$ among the remaining parameters $k$. 
\begin{figure}
\includegraphics[width=\sizefactor \linewidth]{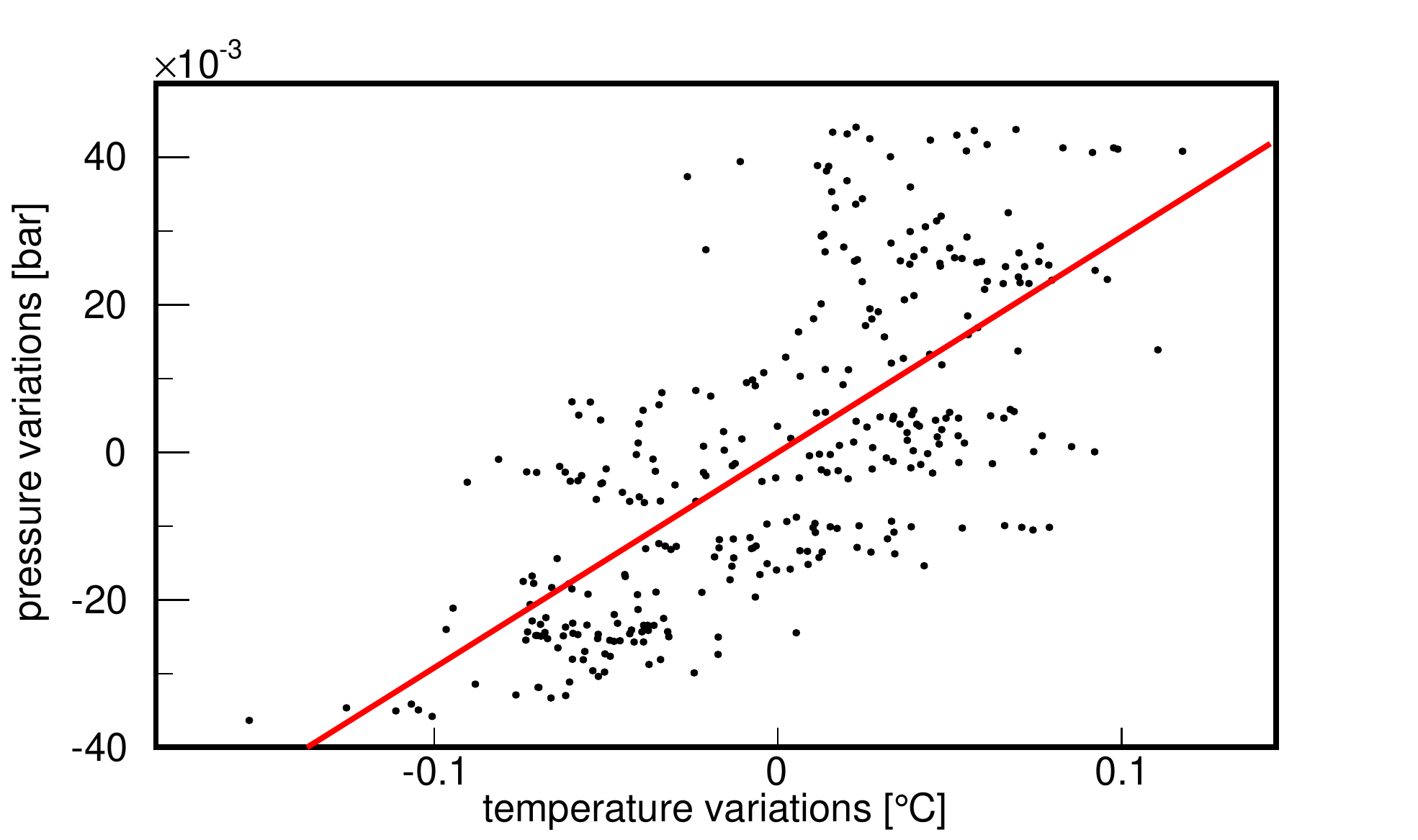}
\caption{Correlation between pressure and temperature in the gas cell. The parameters shown already have the correlation with the laser asymmetry removed.
The error on the data points are about $1\times 10^{-2}$~$^\circ$C and 1~mbar, indicating that other environmental parameters play a role in the data scatter.}\label{fig:PTgas}
\end{figure}
An example is the correlation between pressure in the gas cell and its temperature, as shown in Fig.~\ref{fig:PTgas}. Of course, a correlation between pressure and temperature is to be expected. To see whether other parameters $j$ are relevant we repeat the procedure with the corrected data.
In Appendix~\ref{appendixB} we give a formal account of the correction procedure and the criteria used.

It is not unlikely that one or more of the parameters has an apparent sidereal dependence. Its amplitude $A_j$ can be determined by fitting $A_j\sin (\Omega t^i +\phi)$ to $p_j^i$.
If $D_j A_j$ is significantly deviating from zero the procedure described above might also partly remove the actual LIV signal, as we will discuss in Section~\ref{sec:sensitivity}.
This correction procedure is a crucial step in the analysis. The laser asymmetry was found to drift with the ambient temperature and would have resulted in a sidereal signal of $D_{las} A_{las} = (-2.0\pm 0.3)\times 10^{-4}$.

Data set III has the highest statistical power. For this data set also less significant corrections for pressure and temperature of the buffer gas were made. The first two data sets with much lower statistics allowed only for the laser power correction.
Similarly we determine the magnitude $D_j A_j$ for the remaining parameters that were not significant enough to give a noticeable effect on $\Delta_\text{LIV}$.
Applying the decorrelation procedure among these remaining parameters we find the magnitudes $D_j A_j$.
The individual values for data set III are shown in Table~\ref{tab:sys}.
We take the sum over remaining parameters $\sigma_{\text{corr}} = \sqrt{\sum_j (D_j A_j)^2}$ as the remaining systematic uncertainty of the correction procedure.

 \begin{table}
\centering
\begin{tabular}{|p{33mm}||c|C{15mm}|C{15mm}|}
\hline
\hline
\multirow{2}{*}{parameter $j$} & \multirow{2}{*}{$D_j / \sigma_{D_j}$} & \multicolumn{2}{c|}{$D_j A_j$ ($\times 10^{-5}$)}\\
\cline{3-4}
&& $\hat{N}^1$ & $\hat{N}^2$ \\
\hline
\hline
$\gamma_2$ temperature&$1.1$	& $-1.3$ &	$0.4$\\
\hline
$\beta_2$ temperature	&$0.68$	&	$-0.5$		& $-1.8$			\\
\hline
$\beta_1$ temperature	&$1.2$	&	 $1.3$		& $0.0$			\\
\hline
laser average 			&$1.1$	&	 $1.2$		& $0.8$			\\
\hline
$\gamma_1$ temperature&$0.96$	&	 $-1.5$		& $-1.0$			\\
\hline
beam intensity			&$0.88$	&	 $0.3$		& $1.3$			\\
\hline
hall temperature		&$0.14$	&	$-0.3$		& $0.0$			\\
\hline
\hline
$\sigma_\text{corr}$		&	& 	$2.7$		& $2.6$			\\
\hline
\hline
\end{tabular}
\caption{\label{tab:sys}Systematic uncertainties for data set III due to the parameters $\Delta_\text{LIV}$ was not corrected for.
The last two columns indicate the correlation $D_j$ multiplied by the sidereal amplitude $A_j$ components in the $N^1$ and $N^2$ direction.
The order of the parameters is a result of the decorrelation procedure where initially the temperature of the $\gamma_2$ detector had the largest $D_j/\sigma_{D_j}$.
$\sigma_\text{corr}$ is the sum of $D_j A_j$ added in quadrature.}
\end{table}
\subsection{Experimental sensitivity}\label{sec:sensitivity}
There are two aspects that affect the sensitivity of the measurement. They are the time dependent depolarization and the possibility of accidental removal of the sidereal signal by the correction procedure as described in the previous section.
\subsubsection{Depolarization dependence}\label{depolarization}
The sensitivity to depolarization is parametrized with $P_\text{eff}$.
 As we showed in Ref.~\cite{NIMPolarization} the polarization of the $^{20}$Na sample can be characterized by a time $\tau_\text{pol}\approx 40$ ms for each particle to become polarized after it enters the gas cell and a depolarization time $\tau_\text{depol}\approx$ 1--4\,s. The latter depends on the gas condition and whether the beam was ``on'' or ``off'' (cf. Fig.~4 of~\cite{NIMPolarization}). Therefore, a Monte Carlo simulation was done to find the effective polarization to be used with Eq.~(\ref{eq:LIVsignal}). The test particles appear in the gas cell with a constant rate untill T=2 s. Upon entering they polarize with a rate $\tau_\text{pol}^{-1}$ and depolarize with $\tau_\text{depol}^{-1}$ resulting in a time dependent polarization $P_n(t)$, cf. Ref.~\cite{NIMPolarization}. The test particles decay with a probability of $[\tau(1 \pm P_n(t)\delta\tau)]^{-1}$, where $\delta\tau$ is chosen appropriately small. The accumulated spectra are fitted with a decay time $\tau(1 \pm P_\text{eff}\delta\tau)$ in the region $\text{T} > 2$\,s, from which $P_\text{eff}$ is obtained. We also determine the dependence of $P_\text{eff}$ on the polarization decay parameter $\tau_\text{depol}$ with these simulations.
We find that the weighted average over the three runs to be $\tau_\text{depol}=1.3 \pm 0.3$\, s for which $P_\text{eff}=(0.79\pm 0.09) P(t=2)$ is a good representation of the data taken where $P(t=2)$ is the value from the actual fit of a particular time slice, as discussed in Section~\ref{sec:deltaLIV}.
 \subsubsection{Impact of the correction for systematic errors}\label{sec:sens}

 The procedure for correcting systematic errors could remove part of an actual LIV signal in $\Delta_\text{LIV}(t)$, thus reducing the experimental sensitivity.
After unblinding the data and measuring the limits on the LIV amplitudes ($\xi N^1,\xi N^2$) we investigated the extent  to which this happens. We refer to Appendix~\ref{sec:formalsensitivity} for the formal aspect. We added small amounts of artificial sidereal variation with amplitudes ($\delta (\xi N^1),\delta (\xi N^2)$) to the data points $\Delta_\text{LIV}^i$. The entire analysis procedure is then rerun and by fitting the sidereal amplitude again it is established which fraction can be recovered. This procedure is repeated for a set of values ($\delta (\xi N^1),\delta (\xi N^2)$) to map between the final solution plane ($\xi N^1$, $\xi N^2$) and the observed solution plane. The set of values ($\delta (\xi N^1),\delta (\xi N^2)$) is such that all points within the $1\,\sigma$ confidence region of the measurement are reached in the observed solution plane.
The final values ($\xi N^1$, $\xi N^2$) are obtained by scaling the measurement with the inverse of the recovered fraction for the $N^1$ and $N^2$ directions separately. The bounds on $\xi N^1$ and $\xi N^2$ increase maximally 49\,\%.
\section{Results and interpretation}
\begin{figure}
\centering
\includegraphics[width = \sizefactor \linewidth]{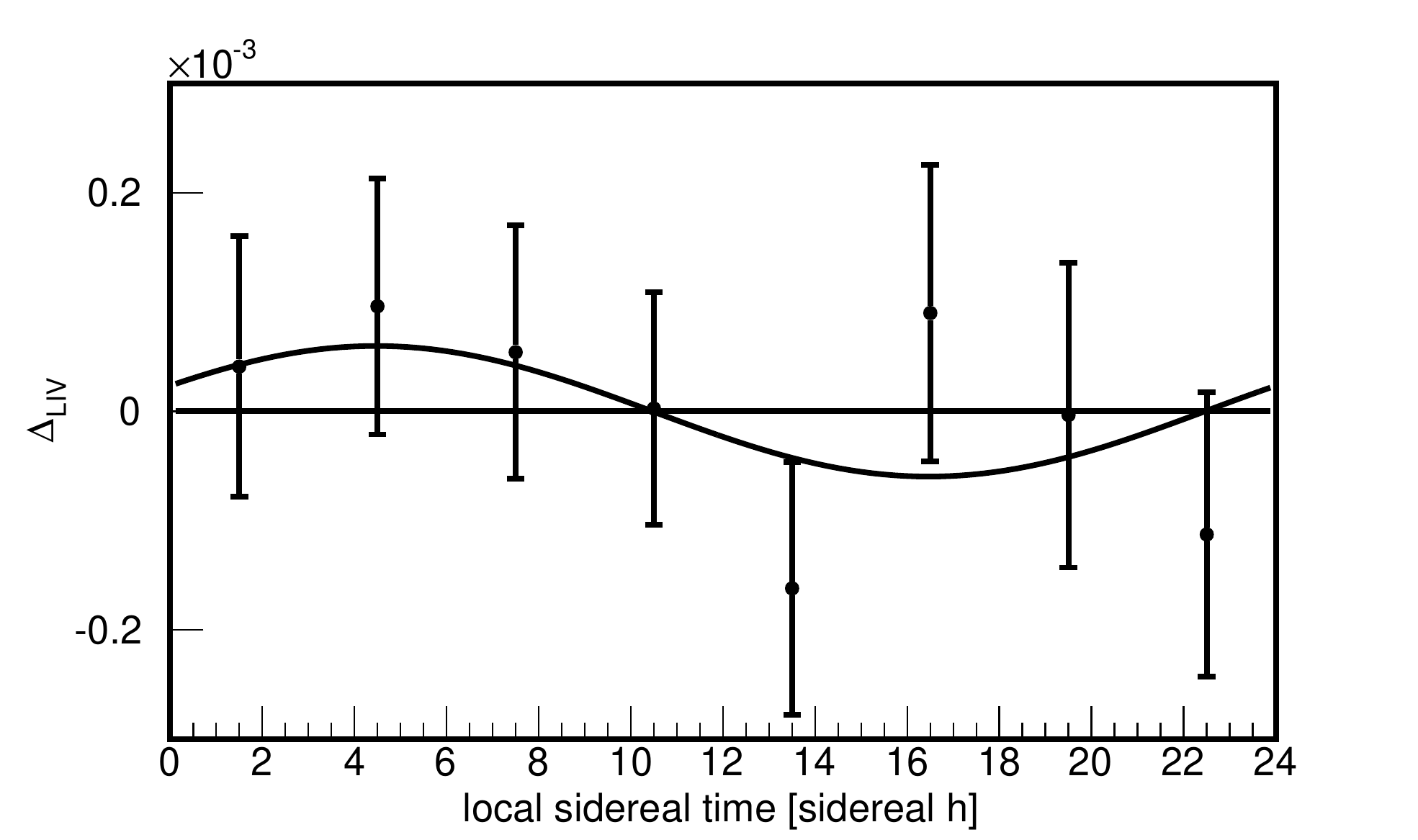}
\caption{The LIV observable $\frac{\Delta \tau}{\tau}\frac{1}{P_{\text{eff}}}$ versus local sidereal time for data set III. For this figure data was binned to 8 bins of 3 sidereal hours width.
Vertical error bars indicate the statistical uncertainty only.
The curved line $y(t) = \xi N^1 \cos\Omega t+\xi N^2\sin\Omega t$ was fitted to the data.
The fit result has $\chi^2/\text{ndf} =  3.2/6$ with $p = 78$\,\%.
Both $\xi N^1$ and $\xi N^2$ are consistent with zero (see Table~\ref{tab:sidampv3}).}
\label{FebLIV}
\end{figure}
$\Delta_{\text{LIV}}$ for data set III as a function of time modulo one sidereal day is shown combined in Fig.~\ref{FebLIV}.
No significant signal for LIV has been found, which yields an upper limit on the LIV amplitude $|A_\text{LIV}|= \xi \sqrt{( N^1)^2+ (N^2)^2}$.
The measurements for data sets I-III are then scaled to correct for the systematic corrections reducing sensitivity (Section~\ref{sec:sens}).
The limits on the LIV amplitudes are shown in Table~\ref{tab:sidampv3} together with the statistical and systematic uncertainties.
The uncertainties in $L$ (Eq.~(\ref{eq:countrate})) are 26\,\%, 10\,\% and 3\,\% for data sets I, II and III, respectively. The correlation between $L$  and the LIV amplitudes $(\xi N^1,\xi N^2)$ leads to an uncertainty in the latter, listed as $\sigma_\text{asymm}$ in Table~\ref{tab:sidampv3}.
It constitutes 8-26\,\% of the total uncertainty for data sets I and II,
while being negligible for data set III.

The analysis considers the cosine and sine components separately. This enabled combining the three data sets while preserving phase information.
Statistical and systematic uncertainties are combined quadratically averaging the measurements I to III to obtain sidereal amplitudes $\xi N^1 = (-0.1 \pm 0.8)\times 10^{-4}$ and $\xi N^2 = (0.2 \pm 1.1)\times 10^{-4}$.
This sets a 90\,\% C.L. limit on sidereal amplitude variations of $|A_\text{LIV}| < 2\times 10^{-4}$.

 \begin{table*}
\centering
 \begin{tabular}{|l|l|C{30mm}|C{30mm}|C{30mm}|}
\hline
\hline
\multirow{2}{15mm}{data set} & \multirow{2}{30mm}{${\left\{\begin{array}{c} \xi N^1 \\ \xi N^2 \end{array}\right.}$~($\times 10^{-4}$)}& \multirow{2}{22mm}{$\sigma_\text{stat}$~($\times 10^{-4}$)} & \multicolumn{2}{c|}{systematic uncertainties} \\
\cline{4-5}
& & & $\sigma_\text{corr}$~($\times 10^{-4}$)& $\sigma_\text{asymm}$~($\times 10^{-4}$)\\
\hline
\hline
I & $\left\{\begin{array}{c} -2.9 \\ -0.8 \end{array}\right.$ & $\begin{array}{c} 4.1 \\ 5.5 \end{array}$ & $\begin{array}{c} 8.8 \\ 5.3 \end{array}$ & $\left.\begin{array}{c} 0.4 \\ 1.6 \end{array}\right.$\\
\hline
II & $\left\{\begin{array}{c} -3.4 \\ -5.2 \end{array}\right.$ & $\begin{array}{c} 3.7 \\ 2.6 \end{array}$ & $\begin{array}{c} 4.3 \\ 3.2 \end{array}$ & $\left.\begin{array}{c} 1.5 \\ 1.3 \end{array}\right.$\\
\hline
III & $\left\{\begin{array}{c} -0.1 \\ 0.6 \end{array}\right.$ & $\begin{array}{c} 0.7 \\ 0.9 \end{array}$ & $\begin{array}{c} 0.4 \\ 0.7 \end{array}$ & $\left.\begin{array}{c} 0.0 \\ 0.0 \end{array}\right.$\\
\hline
\hline
 average & $\left\{\begin{array}{c} -0.1 \\ 0.2 \end{array}\right.$ & \multicolumn{3}{c|}{$\sigma_\text{avg}$~($\times 10^{-4}$) \hspace{2mm} $\left.\begin{array}{c} 0.8 \\ 1.1 \end{array}\right.$} \\
\hline
\hline
\end{tabular}
\caption{\label{tab:sidampv3}Limits on the sidereal amplitudes for Lorentz-invariance violation. The standard deviations refer to the statistical and two systematic uncertainties described in the text.}
\end{table*}

This result is interpreted within the framework that was developed in Ref.~\cite{JN2013}.
The $W$-boson propagator is modified by adding a general Lorentz-invariance violating tensor $\chi^{\mu\nu}$ to the metric tensor.
Evaluating the V-A theory for $\beta$ decay with this modification, observables can be found that break Lorentz (and possibly $CPT$) invariance.
In this framework the relative lifetime difference in Eq.~(\ref{eq:LIVsignal}) is given by $\Delta_{\text{LIV}} = - A_{\text{Wu}}\, \vec{\tilde{\chi}}_i \cdot \hat{J}$.
The vector $\vec{\tilde{\chi}}_i$ is defined in the laboratory frame with components $\tilde{\chi}_i^k \equiv \Im({\epsilon_{klm} \chi^{lm}})$, where $k,l,m$ are the spatial indices and the subscript $i$ labels the imaginary part.
Our experiment is exclusively sensitive to imaginary parts of $\chi$.

To make the result independent of the lab frame we transform the tensor $\chi$ to the sun-centered frame of~\cite{tests}: $\vec{\tilde{\chi}}_i\rightarrow\vec{\tilde{X}}_i$. 
For the present setup with east-west polarization the limit on $|A_\text{LIV}|$ results in a limit $[(\tilde{X}_i^X)^2 + (\tilde{X}_i^Y)^2]^{1/2} < 6\times 10^{-4}$.
Superscripts $X,Y,Z$ refer to the spatial coordinates in the sun-centered frame.
This limit is obtained by 90\,\% coverage of a two-dimensional Gaussian, $|\tilde{X}_i^X|$, $|\tilde{X}_i^Y| < 4\times 10^{-4}$.

We interpret the result in the SME where $\chi^{\mu\nu*}=\chi^{\nu\mu}$ up to order $1/M_W^2$ for $\beta$-decay ($M_W$ is the $W$-boson mass)~\cite{JN2013}.
Using $\tilde{X}_i^X=2X_i^{YZ}$ and $\tilde{X}_i^Y=2X_i^{ZX}$ we find the 90\,\% confidence limits $|X_i^{YZ}|$, $|X_i^{XZ}| < 2\times 10^{-4}$. 
The various limits with references to their definitions are summarized in Table~\ref{tab:xi}.
\section{Conclusions and discussion}
No significant polarization-dependent LIV in the decay rate of $^{20}\text{Na}$ was found at $2\times 10^{-4}$ (90\,\% C.L.).

Bounds on the LIV coefficients from other $\beta$-decay experiments have been expressed in the theoretical frameworks of Refs.~\cite{JN2013} and~\cite{tests}.
Very strong limits were derived~\cite{forbidden} from experiments~\cite{NewmanWiesner,Ullman} searching for an anisotropy of ``forbidden'' $\beta$ decays.
Limits on combinations of real and imaginary coefficients of $\chi$ of order $10^{-8}$ were found.
In absence of fine tuning, these strong limits apply also to the coefficients measured in this work.
The present bounds are, however, uniquely linked to the imaginary part of $\chi$, avoiding possible cancellations of coefficients by fine tuning.
Combining Eq.~(8c) of~\cite{forbidden} with limits from pion data~\cite{BA2013} reduces but does not eliminate the possibility of fine tuning.

With the present method further reduction of the LIV limit could be obtained by higher polarization, higher particle yields, and a longer measurement time.
Use of segmented detectors reduces coincident summing and, therefore,
 reduces systematic effects related to positron contamination of the $\gamma$ signal.
Intense particle sources could be provided by advanced radioactive-beam facilities.

Alternative methods to obtain direct limits on LIV parameters in the weak interaction are very well possible.
A discussion of possible measurements  is given in~\cite{KV2015a}.
For experiments exploiting orbital electron capture a discussion is in~\cite{KV2015}.
We note that there are as yet no values for $\chi_i^{0l}$. This requires to measure the coincidence rate of $\gamma$ particles, and $\beta$ particles perpendicular to the polarization axis in a setup similar to the present one.

\begin{table}
  \begin{center}
    \begin{tabular}{|c| c| r| }
      \hline
      \hline
      Description & Coefficient & {90\,\% C.L.} \\
      \hline
      \hline
      {\footnotesize sidereal variation} & $|A_{\text{LIV}}|$ & \\
       $\frac{\Delta \tau}{\tau}\frac{1}{P_{\text{eff}}}$ & this work & $<2\times 10^{-4}$\\
       & previous work~\cite{SM2013} & $<3\times 10^{-3}$ \\
      \hline
      $\chi$ tensor~\cite{JN2013} & $|\tilde{X}_i^X|$, $|\tilde{X}_i^Y|$ & $<4\times 10^{-4}$\\
      \hline
      SME~\cite{tests,JN2013} & $|X_i^{XZ}|=\left|(k_{\phi\phi}^A)^{XZ}+\frac{1}{2g}(k_{\phi W})^{XZ}\right|$& $<2\times 10^{-4}$\\
       & $|X_i^{YZ}|=\left|(k_{\phi\phi}^A)^{YZ}+\frac{1}{2g}(k_{\phi W})^{YZ}\right|$ & \\
      \hline
      \hline
    \end{tabular}
  \caption{Limits on sidereal amplitudes of $\Delta_{\text{LIV}}$ at 90\,\% C.L. and the corresponding limits for the
$\chi$ tensor formalism and the SME parameters in the sun-centered frame.}\label{tab:xi}
  \end{center}
\end{table}
\section*{Acknowledgments}
We thank the AGOR cyclotron staff for providing the beam and L. Huisman for technical support.
This research was financially supported by the “Stichting voor Fundamenteel
Onderzoek der Materie (FOM)” under Programme 114 (TRI$\mu$P) and ``FOM projectruimte''
08PR2636-1.
\appendix
\section{Formula for the $\gamma$-detector rate for repeated data cycles}\label{appendixA}
While the sodium is introduced to the buffer-gas cell it will decay.
The basic decay rate given in Section~\ref{sec:deltaLIV} is modified by consecutive periods (duration $P = 4.1$\,s) of beam ``on'' (duration $T = 2$\,s) and beam ``off'' (duration $T' = 2.1$\,s) so that
\begin{align}\label{eq:rloop}
 R(t) =& A(1-e^{-T/\tau})e^{-t/\tau} \times \notag\\
&\left[\frac{e^{-T' / \tau}}{1-e^{-P/\tau}} + \begin{cases}
                               \frac{e^{t/\tau} - 1}{1 - e^{-T/\tau}} & \mbox{if $t < T$} \\
				e^{T/\tau} & \mbox{if $t \geq T$}
                              \end{cases}\right]\,
\end{align}
with $A$ the scaling parameter and $\tau$ the $^{20}\text{Na}$ lifetime.
The first term describes decay of sodium nuclei remaining from the previous periods which was approximated with an infinite sequence.
If there were only $N$ previous periods the error of this approximation is $e^{-(N+1)P/\tau}$.
\section{}
\subsection{Procedure used for systematic correction}\label{appendixB}
\newcommand{\LL}{{\bf\Delta}}
This appendix describes the correction procedure to correct data points $\Delta_\text{LIV}^i$ for a correlation with a separately measured parameter $j$ using its value $p_j^i$. We use the vector notation for the set $\{i\}$ and drop the subscript ``$\text{LIV}$''. We start with the following Ansatz:
\begin{align}\label{delta}
\LL=\LL_{R}+\sum_j D_j {\bf p}_{j}\;,
\end{align}
where $\LL$ are the measured values for each slice and $\LL_{R}$ is the true LIV signal. The sum contains contributions from parameters $j$ that have an impact of $D_j$ on the value of the LIV signal but that could not be modeled in the fit of the time-dependent $\gamma$ rates. 
The data are renormalized so that 
\begin{align}\label{offset}
\langle \LL\rangle=\langle{\bf p}_j\rangle=0\;.
\end{align}
The correlation $(\LL, {\bf p}_j)$ is
\begin{align}\label{startexpression}
D'_j=\frac{\langle{\bf p}_j\cdot \LL\rangle}{\langle {\bf p}_j^2\rangle}\;.
\end{align}
Inserting Eq.~(\ref{delta}) into Eq.~(\ref{startexpression}) one finds
\begin{align}
D'_j=D_j +\frac{\langle {\bf p}_j \cdot \LL_R\rangle +\sum_{j\neq k} D_j\langle {\bf p}_k\cdot{\bf p}_j\rangle}{\langle {\bf p}^2_j\rangle }\;.
\end{align}
Correcting ${\bf \Delta}$ for parameter $j$ one finds
\begin{align}\label{eq:Lprime}
 \LL\rightarrow \LL'=\LL-D'_j {\bf p}_j=\LL_R -\frac{\langle {\bf p}_j\cdot \LL_R\rangle +\sum_{k\neq j}D_j\langle{\bf p}_k \cdot{\bf p}_j\rangle}{\langle {\bf p}^2_j\rangle }{\bf p}_j+\sum_{k\neq j}D_k{\bf p}_k\;.
\end{align} Therefore, the correction is successful, when there is no residual correlation $\langle {\bf p}_j\cdot \LL_R\rangle$ or $\langle{\bf p}_k \cdot{\bf p}_j\rangle$.

It is clear that if there are several parameters that are driven by a common parameter, \textit{e.g.} a temperature, we cannot use the raw values of $p_k$. Instead one should remove the correlations so that $\langle p_k p_l\rangle=0$. This can be done analogously 
\begin{align}\label{eq:pprime}
{\bf p'}_k={\bf p_k}-\frac{\langle {\bf p}_k \cdot {\bf p}_j\rangle}{\langle {\bf p}^2_j\rangle }{\bf p}_j\;.
\end{align} 
If $\langle {\bf p}_j\cdot \LL_R\rangle\neq 0$ it appears that one could remove the actual LIV signal. This aspect cannot be studied with blinded data. After unblinding the data one can add an artificial amount of sidereal signal and determine to what extent it is removed in the correction procedure. This is then taken into account in the final result as a reduced sensitivity. This is discussed in the next section of this appendix.

We have not yet considered the statistical uncertainty of the data for clarity of argument. Here we modify the expressions above to include the uncertainty analysis. In this case the coefficients $D_j$ are obtained in a least-squares procedure using the errors $\bf \sigma$ in $\LL$. Using the full notation one has 
\begin{align}
D_j=\frac{\sum_i\frac{\Delta^i_\text{LIV}p_j^i}{(\sigma^i)^2}}{\sum_i(\frac{p_j^i}{\sigma^i})^2}\; \text{\ \   with error\ \ } \sigma_{D_j}=\frac{1}{\frac{1}{N}\sum(\frac{p_j^i}{\sigma^i})^2}\;.
\end{align} 
These expressions require as in Eq.~(\ref{offset}) that $\sum\frac{\Delta_\text{LIV}^i}{(\sigma^i)^2}=0$, which is the case by definition, and although the $p_j^i$ are normalized using $\sum p_j^i=0$ also $\sum(\frac{p_j^i}{\sigma^i})^2\approx 0$.

The order in which the corrections on $\bf\Delta$ as in Eq.~(\ref{eq:Lprime}) for its dependence on parameter $j $ are done is based on the significance $D_j/\sigma_{D_j}$.
First the most significant correction is taken, using the parameter $j$ with the largest value $D_j/\sigma_{D_j}$. Also all parameters $k, (k\neq j)$ are corrected as in Eq.~(\ref{eq:pprime}). The modified $\Delta_\text{LIV}$ and $p_j$ are then used to search the next most significant contribution. This procedure is repeated with the remaining parameters until there is no parameter left with $D_j/\sigma_{D_j}>3$. This cutoff is used to avoid overcorrection on statistically insignificant dependencies.

The errors in $\LL$ are propagated as
\begin{align}
\boldsymbol{\sigma}'_\text{LIV}=\sqrt{{\boldsymbol\sigma}^2_\text{LIV}+(\sigma_{D_j}{\mathbf{p}_j})^2+(D_j{\boldsymbol\sigma}_{p_j})^2}\;.
\end{align}
From this it can be seen that only parameters $j$ with significant values of $D_j$ should be considered in the correction procedure.

The strong selection on the parameters that are considered for a correction of the LIV signal may mean that corrections are incomplete. For this we consider the set of parameters $j'$ that were not used in the correction procedure and determine their maximal contribution to a sidereal amplitude ignoring the phase, which is $D_{j'}A_{j'}$. This allow us to access the systematic error in our procedure by selecting again the largest contribution $D_{j'}/\sigma_{D_{j'}}$, applying the correction of Eq.~(\ref{eq:pprime}) and observing the convergence of these errors to a common noise level.
\subsection{The sensitivity factor}\label{sec:formalsensitivity}
To find to which extent a true $\LL_R$ could be removed due to a finite value of $\langle {\bf p}_j\cdot \LL_R\rangle$ in Eq.~(\ref{eq:Lprime}) consider the following. One can write
\begin{align}
{\bf p}_j=A_j \LL_R + {\bf s}\;,
\end{align}
where the parameter has a time dependence identical to the sidereal frequency of $\LL_R$ with magnitude $A_j$ and we assume a contribution $ {\bf s}$ that is effectively stochastic. The amount removed from $\LL_R$ is then
\begin{align}\frac{\langle A_j\LL_R^2+{\bf s}\LL_R\rangle}{\langle A_j^2\LL_R^2 +{\bf s}^2 + 2{\bf s}\LL_R\rangle}(A_j \LL_R + {\bf s})\;.
\end{align}
Therefore, if the parameter has no stochastic part, \textit{i.e.} ${\bf s}=0$, the entire signal will removed. However, in the more usual case $\langle {\bf s}^2\rangle \gg \langle A_j^2\LL_R^2\rangle$, a much smaller fraction of $\LL_R$ is lost \textit{i.e.}\ $\langle A_j^2\LL_R^2\rangle/\langle {\bf s}^2\rangle$.
In this work the sensitivity factor is obtained by adding an artificial sidereal dependence and determining how much of this added signal survives the correction procedure; this is defined as the sensitivity factor.
\bibliographystyle{apsrev4-1}
\providecommand{\noopsort}[1]{}\providecommand{\singleletter}[1]{#1}%
\end{document}